\documentclass[aps,prb,twocolumn,showpacs]{revtex4}
\usepackage{graphicx}
\usepackage{amsmath}
\usepackage{amssymb}

\begin{document}

\title{Magnetic properties of two-dimensional charged spin-1 Bose gases}

\author{Yingxue Chen}
\author{Jihong Qin}\email[Corresponding author. {\it E-mail address}: ]{jhqin@sas.ustb.edu.cn}
\author{Qiang Gu}
\affiliation{Department of Physics, University of Science and
Technology Beijing, Beijing 100083, China}

\begin{abstract}
Within the mean-field theory, we investigate the magnetic properties
of a charged spin-1 Bose gas in two dimension. In this system the
diamagnetism competes with paramagnetism, where Lande-factor $g$ is
introduced to describe the strength of the paramagnetic effect. The
system presents a crossover from diamagnetism to paramagnetism with
the increasing of Lande-factor. The critical value of the
Lande-factor, $g_{c}$, is discussed as a function of the temperature
and magnetic field. We get the same value of $g_{c}$ both in the low
temperature and strong magnetic field limit. Our results also show
that in very weak magnetic field no condensation happens in the two
dimensional charged spin-1 Bose gas.



\end{abstract}

\maketitle

\section{Introduction}
The two-dimensional (2D) electronic systems such as electronic
states of semiconductor surface and interface have been considerably
studied in solid-state physics for a long time \cite{Ando}. On the
other hand, less attention has been taken to the magnetism of Bose
gases. The charged Bose gas has played an important role in studying
the conventional superconductivity. Schafroth \cite{Schafroth}
indicated that a three-dimensional (3D) charged Bose gas can exhibit
a Meissner-Ochsenfeld effect at a low temperature. It is well known
that the Bardeen-Cooper-Schrieffer (BCS) theory \cite{Bardeen} is
successful in describing the conventional superconductor. As for
real-space pairs, charged Bose gas can still be used to understand
the magnetism of superconductivity. Apart from the 3D charged Bose
gas, the charged Bose gas in two dimension also deserves attention.
May \cite{May1} demonstrated the possibility of the occurrence of a
Meissner-Ochsenfeld effect in the 2D charged Bose gas. Although the
Meissner effect is imperfect in the 2D case, the diamagnetism is
extremely large. Recently, charged real-space bosons has been used
to explain the diamagnetism in the normal state of high temperature
cuprate superconductors \cite{Alexandrov1}. The 2D $\text{CuO}_{2}$
plane is an important common feature for the doped cuprate
superconductors \cite{Kastner}, and it seems evident that this plane
dominates the nonconventional behaviors. While the 2D charged Bose
gas may act as a model for understanding this superconductivity
\cite{Alexandrov2}. Accordingly, besides the interest of Bose gases
in their own right, the discovery of high temperature
superconductivity also stimulates the renewed research interest in
the 2D charged Bose gas.

Theoretically, the dielectric response of the 2D charged Bose gas
has been investigated both in zero magnetic field \cite{Hines} and
in nonzero magnetic field \cite{Bardos}. Davoudi {\it et al.}
studied the ground-state properties of the 2D charged Bose gas with
considering the system interacting via a logarithmic potential
\cite{Davoudi}. As far as the magnetic properties of an ideal 2D
charged Bose gas \cite{May1,Daicic1,Zyl} is concerned, the charge
degree has been focused on, while the spin degree of freedom is not
considered. It is known that the charge degree of freedom in a
magnetic field induces the diamagnetism, while the paramagnetism is
due to the spin degree of freedom. Both the diamagnetism and
paramagnetism are important equally for the magnetism of charged
spinor Bose gas. Therefore it will be an interesting issue to study
the interplay of the diamagnetism and paramagnetism in the 2D
charged spinor Bose gas.

The magnetic properties of the charged Bose gas have been discussed
based on different methods \cite{Daicic2,Toms}. In a previous paper
\cite{Jian} we have discussed the magnetic properties of the charged
spin-1 Bose gas in three dimension. Furthermore, the ferromagnetic
phase transition has also been investigated in the 3D charged spin-1
Bose gases with ferromagnetic coupling \cite{Qin}. In this paper, we
will focus on the magnetism of charged spin-1 Bose gas in two
dimension. We show that no evidence of Bose-Einstein condensation
(BEC) is seen for the 2D charged spin-1 Bose gas in the very weak
magnetic field, which is a significant difference between the 2D and
3D charged Bose gases \cite{May2}. In section II, a model consisting
of both the Landau diamagnetism and Pauli paramagnetism for the 2D
charged spin-1 Bose gas is constructed. The magnetization density is
calculated by analytical derivation. In section III, the results are
obtained and the corresponding discussions are given. Besides of the
numerical results, the analytical solution at the limit of high
temperature and weak magnetic field is also presented. In section
IV, we give a summary.

\section{The Model}

The Landau levels quantized from the orbital motion of the 2D
charged bosons with charge $q$ and mass $m^{\ast}$ in the effective
magnetic field $B$ are
\begin{eqnarray}\label{Deg}
\epsilon^l_j=\left(j+\frac{1}{2}\right)\hbar\omega,
\end{eqnarray}
where $j=0,1,2,\ldots$ labels different Landau levels and the
gyromagnetic frequency $ \omega=qB/(m^{\ast}c)$. The degeneracy of
each Landau level is
\begin{eqnarray}\label{dege}
D_L=\frac{q B S}{2\pi\hbar c},
\end{eqnarray}
where $S$ is the section area of x-y plane of the system. For a
spin-1 boson, the intrinsic magnetic moment related with the spin
degree of freedom induces the Zeeman energy levels split in the
magnetic field,
\begin{eqnarray}\label{Param}
\epsilon^{ze}_\sigma=-g\frac{\hbar q}{m^\ast c}\sigma
B=-g\sigma\hbar\omega,
\end{eqnarray}
where $g$ is the Lande-factor and $\sigma$ denotes the spin-z index
of Zeeman state $\left| {F=1,m_F=\sigma} \right\rangle$ ($\sigma= 1,
0, -1$). Therefore the effective Hamiltonian can be constructed as,
\begin{eqnarray}\label{Hamilt}
\bar{H}-\mu{N}=D_L\sum_{j,\sigma}\left(\epsilon^l_j+\epsilon^{ze}_\sigma-\mu\right)n_{j\sigma},
\end{eqnarray}
where $\mu$ is the chemical potential. Then we obtain the grand
thermodynamic potential,
\begin{align}\label{T1}
\Omega_{T\neq0}&=-\frac{1}{\beta}\ln\mathrm{Tr}e^{-\beta(\bar{H}-\mu{N})}
\nonumber \\
&=\frac{1}{\beta}D_L\sum_{j,\sigma}\ln\left[1-e^{-\beta\left(\epsilon^l_j+\epsilon^{ze}_\sigma-\mu\right)}\right]
\end{align}
where $\beta=1/({k_BT})$. Through Taylor expansion, equation
(\ref{T1}) can be evaluated as
\begin{align}\label{T2}
\Omega_{T\neq0}=-\frac{m^\ast\omega{S}}{2\pi\hbar\beta}\sum^{\infty}_{l=1}\sum_{\sigma}{\frac{l^{-1}e^{-l\beta\left(\frac{1}{2}\hbar\omega-g\sigma\hbar\omega-\mu\right)}}{1-e^{-l\beta\hbar\omega}}}
\end{align}
Some compact notation for the class of sums is introduced for simplicity,
\begin{align}\label{sum}
B^\sigma_\kappa[\alpha,\delta]=\sum^{\infty}_{l=1}{\frac{l^{\alpha/2}e^{-l\beta\hbar\omega\left(\frac{1}{2}-g\sigma-\frac{\mu}{\hbar\omega}+\delta\right)}}{{\left(1-e^{-l\beta\hbar\omega}\right)}^\kappa}}
\end{align}
With this notation, equation (\ref{T2}) can be rewritten as
\begin{align}\label{T3}
\Omega_{T\neq0}=-\frac{m^\ast\omega{S}}{2\pi\hbar\beta}\sum_{\sigma}B^\sigma_1[-2,0]
\end{align}
Then the density of the 2D bosons $n=N/S$ can be obtained through
the grand thermodynamic potential,
\begin{align}\label{dens}
n&=-\frac{1}{S}\left(\frac{\partial\Omega_{T\neq0}}{\partial{\mu}}\right)_{T,S}
\nonumber\\
&=\frac{m^\ast\omega}{2\pi\hbar}\sum_{\sigma}B^\sigma_1[0,0]
\end{align}
The magnetization density can be derived from the grand
thermodynamic potential,
\begin{align}\label{magd1}
M_{T\neq0}&=-\frac{1}{S}\left(\frac{\partial\Omega_{T\neq0}}{\partial{B}}\right)_{T,S}
\nonumber\\
&=\frac{q}{2\pi\hbar\beta c}\sum_{\sigma} \bigg\{B^\sigma_1[-2,0]+\beta\hbar\omega
\nonumber\\
&\times\left[
\left(g\sigma-\frac{1}{2}\right)B^\sigma_1[0,0]-B^\sigma_2[0,1]
\right] \bigg\}
\end{align}
Hereafter we introduce some dimensionless parameters for
computational convenience, such as $\bar M={m^\ast c M}/({n \hbar
q}), \bar\omega={\hbar\omega}/({k_BT^\ast}),
\bar\mu={\mu}/({k_BT^\ast}), t={T}/{T^\ast}$, with the
characteristic temperature $T^\ast$ is determined by
$k_BT^\ast={2\pi\hbar^2 n}/{m^\ast}$. Thus the equation (\ref{dens})
and (\ref{magd1}) can be further expressed as
\begin{align}\label{selfeq}
1=\bar{\omega}\sum_\sigma\bar B^\sigma_1[0,0]
\end{align}
\begin{align}\label{magd2}
&\bar M_{T\neq0}=t\sum_\sigma \bigg\{ \bar B^\sigma_1[-2,0]+\frac{\bar\omega}{t}
\nonumber\\
&\times\left[ \left(g\sigma-\frac{1}{2}\right)\bar
B^\sigma_1[0,0]-\bar B^\sigma_2[0,1] \right] \bigg\}
\end{align}
respectively, where
\begin{align}
\bar
B^\sigma_\kappa[\alpha,\delta]=\sum^{\infty}_{l=1}{\frac{l^{\alpha/2}e^{-l\frac{\bar\omega}{t}\left(\frac{1}{2}-g\sigma-\frac{\bar\mu}{\bar\omega}+\delta\right)}}{{\left(1-e^{-l\frac{\bar\omega}{t}}\right)}^\kappa}}
\end{align}
and $\bar\mu$ is the dimensionless parameter of the chemical
potential, which can be determined from the mean-field
self-consistent equation (\ref{selfeq}).

\section{Results and discussions}

\begin {figure}[t]
\center\includegraphics[width=0.45\textwidth,keepaspectratio=true]{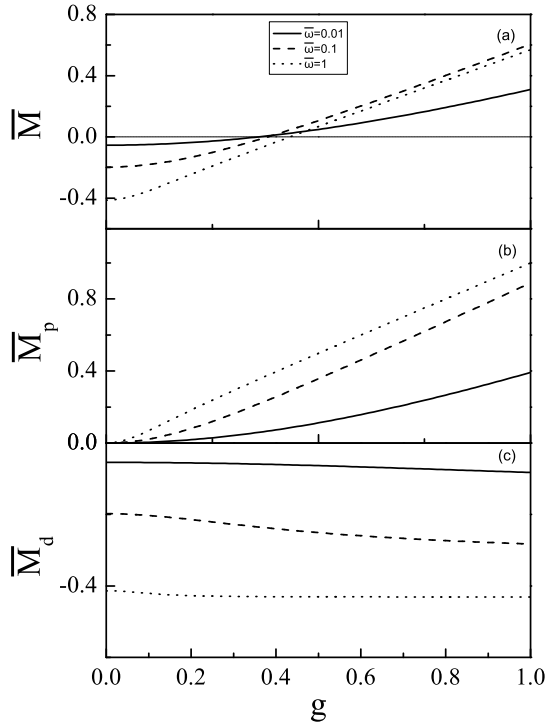}
\caption{(a) The total magnetization density ($\bar M $), (b) the
paramagnetization density ($\bar {M_p} $), and (c) the
diamagnetization density ($\bar {M_d} $) as a function of
Lande-factor $g$ at $t=0.1$. The field is chosen as
$\bar\omega$=0.01 (solid line), 0.1 (dashed line), and 1 (dotted
line). \label{fig1}}
\end{figure}

In the following calculations we will discuss the competition
between paramagnetism and diamagnetism in the 2D charged spin-1 Bose
gases. Meanwhile a comparison with the 3D results \cite{Jian} will
also been presented.

Firstly, we look at the evolution of total magnetization density
$\bar M$ with the Lande-factor $g$. As shown in Fig. 1(a), $\bar M$
is negative when $g$ is small, and changes gradually to positive
after $g_{c}$. Here $g_{c}$ is the critical value of the
Lande-factor $g$, where $\bar M$ changes its sign from negative to
positive. In our model, $\bar M$ contains contributions come from
both paramagnetism and diamagnetism. Fig. 1(b) indicates that the
paramagnetization density $\bar M_{p}$ grows monotonously with $g$
for fixed $\bar \omega$. As shown in Fig. 1(c), the diamagnetization
density $\bar M_{d}=\bar M-\bar M_{p}$ strengthens a little with
increasing $g$ until saturates in the large $g$ region. Comparing to
the 3D case, the diamagnetization density of 2D system is slightly
dependent on $g$. While in the 3D case the diamagnetization density
increases quickly with increasing $g$, especially in the small $g$
and weak magnetic field region. The curves of different magnetic
field also show the stronger magnetic field is, the stronger
paramagnetism and diamagnetism are at the same $g$. However the
evolutionary tendency is similar qualitatively for each fixed
magnetic field. According to the curves above, it is obviously that
the total magnetization density is the result of the competition
between paramagnetism and diamagnetism. For the case of small $g$,
diamagnetism plays a major role. But the diamagnetism is covered up
by the paramagnetism when $g$ exceeds $g_{c}$.

\begin {figure}[t]
\center\includegraphics[width=0.45\textwidth,keepaspectratio=true]{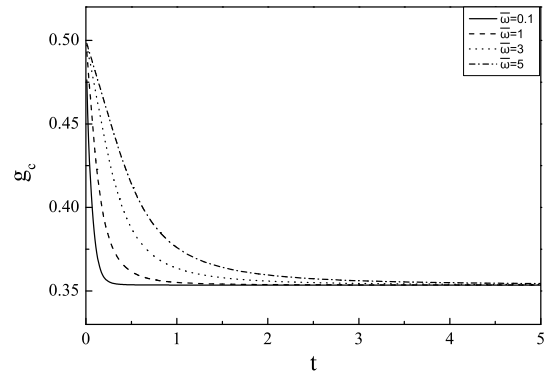}
\caption{The critical value of Lande-factor, $g_c$, as a function of
temperature $t$ with fixed magnetic field $\bar\omega$, where
$\bar\omega=$0.1 (solid line), 1 (dashed line), 3 (dotted line), and
5 (dash-dotted line). \label{fig2}}
\end{figure}

\begin {figure}[t]
\center\includegraphics[width=0.45\textwidth,keepaspectratio=true]{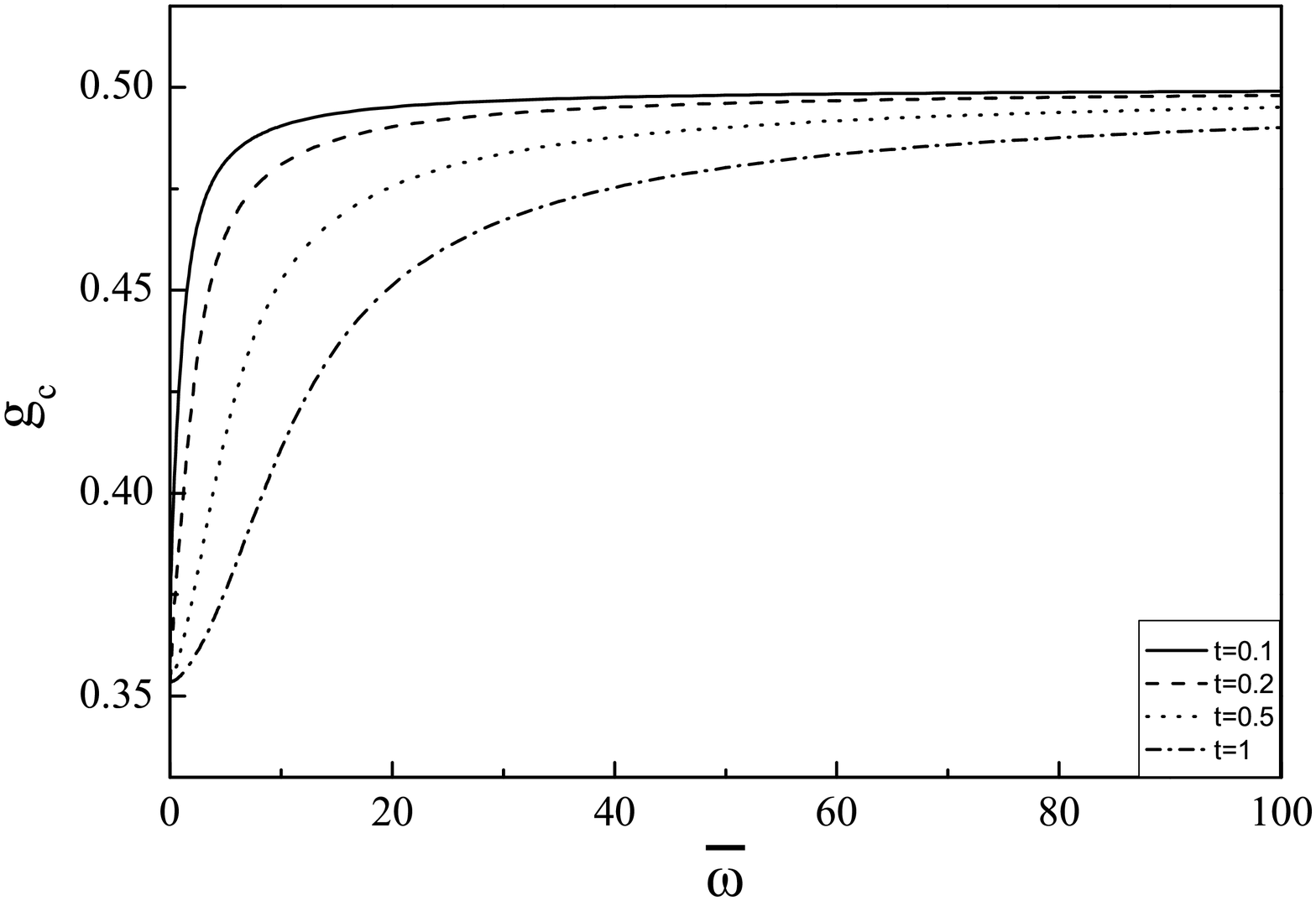}
\caption{The critical value of Lande-factor, $g_c$, as a function of
magnetic field $\bar\omega$ with fixed temperature $t$, where
$t=$0.1 (solid line), 0.2 (dashed line), 0.5 (dotted line), and 1
(dash-dotted line).\label{fig3}}
\end{figure}

The above results show that the shift point of Lande-factor plays a
significant role in describing the competition between the
paramagnetism and diamagnetism. Therefore we will focus on $g_{c}$
hereafter. $g_{c}$ can be obtained by setting $\bar M$=0 in equation
(\ref{magd2}). It indicates that $g_{c}$ is dependent of temperature
and magnetic field. Fig. 2 plots the dependence of $g_{c}$ with
temperature for fixed $\bar \omega$. In the low temperature limit,
$g_{c}$ tends to 0.5, and $g_{c} \approx 0.35$ in the high
temperature limit. In the low temperature region, the smaller $\bar
\omega$ is, the faster $g_{c}$ declines. To further understand the
effect of the magnetic field on $g_{c}$, $g_{c}$ as a function of
magnetic field with fixed temperature is plotted in Fig. 3. It is
shown that the $g_{c}$ limit in the weak magnetic field resembles
that in the high temperature. It is interesting that $g_{c}$ reaches
a uniform value when the temperature $t \longrightarrow \infty$ or
the magnetic field $\bar \omega \longrightarrow 0$. Now we do some
analytical solution to verify our numerical results.

\begin {figure}[t]
\center\includegraphics[width=0.45\textwidth,keepaspectratio=true]{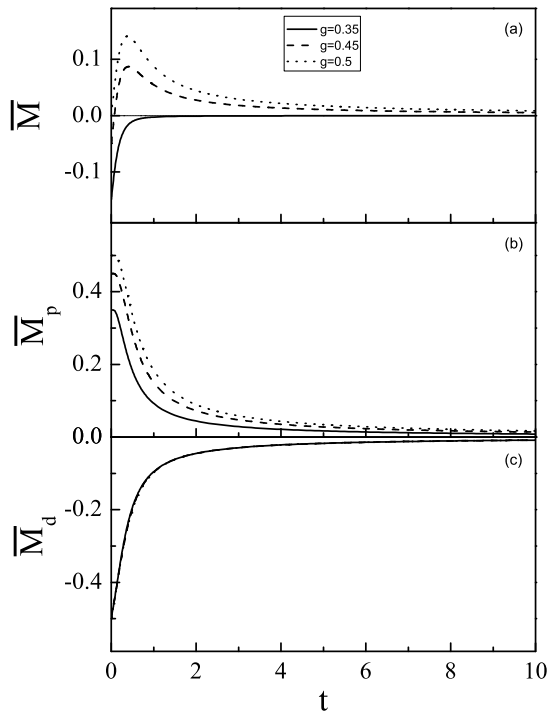}
\caption{(a) The total magnetization density ($\bar M $), (b) the
paramagnetization density ($\bar {M_p} $), and (c) the
diamagnetization density ($\bar {M_d} $) as a function of
temperature $t$ at $\bar\omega=1$, where Lande-factor $g$=0.35
(solid line), 0.45 (dashed line), and 0.5 (dotted line).
\label{fig4}}
\end{figure}

In the high temperature limit, Bose-Einstein statistics reduces to
Maxwell-Boltzmann statistics. This makes it possible to get an exact
value of $g_{c}$. Within Maxwell-Boltzmann statistics, the grand
thermodynamic potential is formally expressed as
\begin{align}\label{ht1}
\Omega_{T\neq0}=-\frac{1}{\beta}D_L\sum_{j,\sigma}e^{-\beta\left(\epsilon^l_j+\epsilon^{ze}_\sigma-\mu\right)}
\end{align}
In this case the dimensionless chemical potential $\bar\mu'$ can be
obtained by
\begin{align}\label{ht2}
1=\bar{\omega}\sum_\sigma\frac{e^{-\frac{\bar\omega}{t}\left(\frac{1}{2}-g\sigma-\frac{\bar\mu'}{\bar\omega}\right)}}{1-e^{-\frac{\bar\omega}{t}}}
\end{align}
And the dimensionless magnetization density based on
Maxwell-Boltzmann statistics can be re-expressed as
\begin{align}\label{ht3}
&\bar M_{T\neq0}^{B}=t\sum_\sigma \Bigg\{
\frac{e^{-\frac{\bar\omega}{t}\left(\frac{1}{2}-g\sigma-\frac{\bar\mu'}{\bar\omega}\right)}}{1-e^{-\frac{\bar\omega}{t}}}
\nonumber\\
&\times\left[1+\frac{\bar\omega}{t}\left(g\sigma-\frac{1}{2}-\frac{e^{-\frac{\bar\omega}{t}}}{1-e^{-\frac{\bar\omega}{t}}}\right)
\right] \Bigg\}
\end{align}

Now we substitute equation (\ref{ht2}) into (\ref{ht3}), then we
obtain
\begin{align}\label{ht4}
\bar M_{T\neq0}^{B}=\frac{1}{x}-\frac{1}{2}-\frac{1}{e^x-1}+g
\frac{e^{gx}-e^{-gx}}{e^{gx}+e^{-gx}+1}
\end{align}
where $x={\bar\omega}/{t}$. As $\bar M_{T\neq0}^{B}=0$, $g_{c}$ can
be resolved from the analytical formula (\ref{ht4}). In $t
\longrightarrow \infty$ or $\bar \omega \longrightarrow 0$,
$g_{c}={1}/{\sqrt{8}}\approx 0.354$. This analytical results is
agreement with that presented in Fig. 2 and Fig. 3.

\begin {figure}[t]
\center\includegraphics[width=0.45\textwidth,keepaspectratio=true]{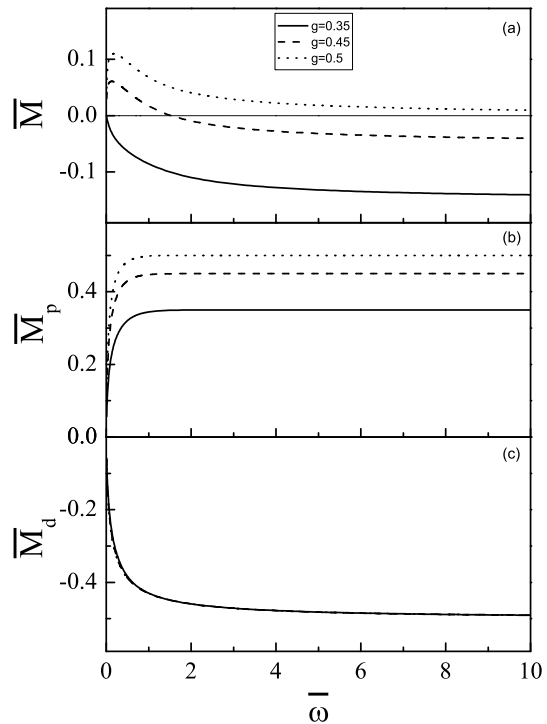}
\caption{(a) The total magnetization density ($\bar M $), (b) the
paramagnetization density ($\bar {M_p} $), and (c) the
diamagnetization density ($\bar {M_d} $) as a function of magnetic
field $\bar\omega$ at $t=0.1$, where Lande-factor $g$=0.35 (solid
line), 0.45 (dashed line), and 0.5 (dotted line). \label{fig5}}
\end{figure}

In order to manifest what affect paramagnetism and diamagnetism
respectively in detail, Fig. 4 and Fig. 5 are plotted. Fig. 4 is
plotted in a definite magnetic field $\bar \omega=1$, while Fig. 5
is given with a fixed temperature $t=0.1$. Since the two limits of
$g_{c}$ are $1/2$ (low temperature or strong magnetic field) and
${1}/{\sqrt{8}}\approx 0.354$ (high temperature or weak magnetic
field), the results will be shown for the three character parameters
of $g$, $g=0.35$, $g=0.45$ and $g=0.5$. We explore the similarity
between Fig. 4(a) and Fig. 5(a). The total magnetization density
$\bar M$ curves of $g=0.35$ always present diamagnetism in spite of
the temperature and magnetic field, while $\bar M$ maintain
paramagnetism when $g=0.5$. But $\bar M$ exhibits a transition
between paramagnetism and diamagnetism at $g=0.45$. This is because
$g=0.45$ locates in the intermediate region between
$g={1}/{\sqrt{8}}$ and $g=1/2$. It can be seen from Fig. 4(c) that
diamagnetism depends mostly on the temperature in the low
temperature region, regardless of the Lande-factor, until saturates
at high temperature. Besides the temperature, the Lande-factor plays
an important role in the paramagnetism, which is shown in Fig. 4(b).
In the weak magnetic field indicated by Fig. 5(c), the magnitude of
magnetic field mostly affects the diamagnetism, in spite of the
Lande-factor. While in Fig. 5(b), Lande-factor dominates the
paramagnetism at the strong magnetic field region.

\begin {figure}[t]
\center\includegraphics[width=0.45\textwidth,keepaspectratio=true]{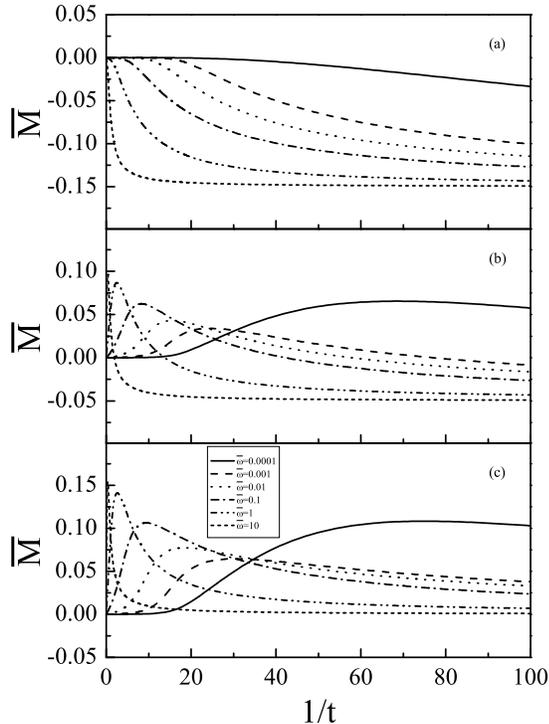}
\caption{Plots of the total magnetization density ($\bar M $) as a
function of $1/t$ at (a) $g$=0.35, (b) $g$=0.45 and (c) $g$=0.5,
with different $\bar\omega$=0.0001 (solid line), 0.001 (dashed
line), 0.01 (dotted line), 0.1 (dash-dotted line), 1
(dash-dot-dotted line), and 10 (short dashed line). \label{fig6}}
\end{figure}

In order to manifest the difference in phase transition between the
2D and 3D charged spin-1 Bose system, the dependence of the total
magnetization density $\bar M$ with $1/t$ at different magnetic
fields is shown in Fig. 6. It is known that the 2D Bose gas with no
external fields does not condense. In Fig. 6(a), with decreasing the
magnetic field, the curve keeps smooth even in very weak magnetic
field, which means no sharp bend appears in the weak field. It
suggests that we can't find a temperature which corresponding to the
BEC temperature in the zero magnetic field. From Figs. 6(b) and
6(c), we can find that no intersection exists at the low temperature
region in the weak magnetic field. This further proves that no
condensation appears with reducing the magnetic field. The results
above is different from our results of the 3D case \cite{Jian}. In
the 3D system, a sharp bend emerges little by little along with
weakening the magnetic field. Therefore a temperature related to the
BEC temperature can be predicted. This proves the key difference
between the 2D and 3D charged spin-1 Bose gas.

\section{Summary}
In summary, we study the competition between paramagnetism and
diamagnetism of a charged spin-1 Bose gas in two dimension based on
the mean-field theory. In a very weak magnetic field, no
condensation is predicted. It indicates the difference between the
2D and 3D system. Despite of this qualitatively distinction, some
magnetic properties in 2D charged spin-1 Bose gas, such as the
critical values of Lande-factor in the high temperature and low
temperature limit, are similar to that of 3D case. The intriguing
behavior may come from the main physics occurring in the 2D plane.
Our results also show in the interplay between paramagnetism and
diamagnetism, Lande-factor plays an important role in the
paramagnetism, while magnetic field impacts significantly on the
diamagnetism.

\acknowledgments
JQ would like to thank Professor Huaiming Guo for
the helpful discussions. This work was supported by the National
Natural Science Foundation of China (Grant No. 11004006), and the
Fundamental Research Funds for the Central Universities of China.

\end{document}